\documentclass[10pt,showpacs,twocolumn]{revtex4}
\usepackage{graphicx}
\usepackage{amsmath}
\usepackage{amssymb}
\usepackage{multirow}
\usepackage{}

\begin{document}
\title{The selection rules of high harmonic generation: the symmetries of molecules and laser fields}
\author{Xi Liu,$^{1}$ Xiaosong Zhu,$^{1}$ \footnote{zhuxiaosong@hust.edu.cn} Liang Li,$^{1}$ Yang Li,$^{1}$ Qingbin Zhang,$^{1}$ Pengfei Lan,$^{1}$ \footnote{pengfeilan@mail.hust.edu.cn} Peixiang Lu$^{1,2}$}

\affiliation{$^1$  School of Physics and Wuhan National Laboratory for Optoelectronics, Huazhong University of
Science and Technology, Wuhan 430074, China\\
$^2$ Laboratory of Optical Information Technology, Wuhan Institute of Technology, Wuhan 430205, China}
\date{\today}

\begin{abstract}
The selection rules of high harmonic generation (HHG) are investigated using three-dimensional time-dependent density functional theory (TDDFT). From the harmonic spectra obtained with various real molecules and different forms of laser fields, several factors that contribute to selection rules are revealed. Extending the targets to stereoscopic molecules, it is shown that the allowed harmonics are dependent on the symmetries of the projections of molecules. For laser fields, the symmetries contributing to the selection rules are discussed according to Lissajous figures and their dynamical directivities. All the phenomena are explained by the symmetry of the full time-dependent Hamiltonian under a combined transformation. We present a systematic study on the selection rules and propose an intuitive method for the judgment of allowed harmonic orders, which can be extended to more complex molecules and various forms of laser pulses.
\end{abstract}  \maketitle

\noindent \section{Introduction}
When atoms or molecules are exposed to intense laser fields, many interesting strong field phenomena will take place \cite{SFP1,SFP2,SFP3,SFP4,SFP5,SFP6}. One of the most attractive phenomena is high harmonic generation (HHG) \cite{HHG1,HHG2,HHG3,HHG4,HHG5}. A typical harmonic spectrum consists of a rapid fall off at first several orders followed by a plateau and a sharp cutoff \cite{cutoff}. The frequencies of yielded harmonics are integer multiples of the driving laser frequency and can reach as high as several hundred harmonic orders \cite{integer}. The HHG has received a large amount of attention in the past decades because it promises two fascinating applications: the HHG provides an effective way to produce coherent extreme ultraviolet attosecond pulse \cite{ASP1,ASP2,ASP3} and it is also a useful tool to gain an insight into electronic structures \cite{ES1,ES2,ES3,ES4} and ultrafast dynamics \cite{Probe1,Probe2,Probe3,Probe4,Probe5} of molecules on the attosecond time scale.

In a linearly polarized (LP) laser field, the harmonic spectrum is composed of only odd harmonics for targets with inversion symmetry, which is due to the interference between adjacent half-cycles. A strict explanation based on symmetries of target-laser configurations can be seen in Ref.\ \cite{BenTal}. For asymmetric molecules, the breaking of inversion symmetry will lead to the emission of even harmonics \cite{Even}. In a circularly polarized (CP) laser field, the molecular harmonic spectrum exhibits rich properties \cite{Circle1,Circle2,Circle3}. It is shown that the allowed harmonic orders for molecular targets driven by CP laser pulses are determined by the discrete rotational symmetries of molecules. Specifically, if a molecule possesses $M$-fold rotational symmetry ($M$ is an integer), the allowed harmonic orders in CP laser field are $kM\pm1$ ($k=0,1,2,...$) \cite{Circ1,Circ2,Circ3}. This selection rule can be explained using the group theory \cite{Circ2}.

In recent years, HHG in counter rotating bicircular (CRB) laser fields has aroused increasing interests \cite{CRB1,CRB2,CRB3,CRB4,CRB5} for its potential to generate circularly polarized extreme ultraviolet radiations \cite{Medisauskas,CirHHG1,CirHHG2,CirHHG3} and for the bicircular high harmonic spectroscopy \cite{Baykusheva}. The CRB laser fields are composed of two coplanar counter rotating CP laser fields with different frequencies. For atomic targets, selection rules of HHG in CRB laser pulse are shown as $kL\pm1$ ($k = 0, 1, 2 \cdots$ ) \cite{atom-sr}, when the CRB laser field possesses $L$-fold rotational symmetry. Recently, the selection rules for molecules in CRB laser fields are discussed \cite{Mauger}. The allowed harmonic orders are $kN\pm1$ ($k = 0, 1, 2 \cdots$ ), where $N$ is the greatest common divisor (GCD) of rotational symmetries of the target and laser field. In the above works, the numerical simulations adopted are restricted to two-dimensional model atoms (and molecular ions) in single active electron (SAE) approximation.

In this paper, we investigate the selection rules of HHG with various real molecules and laser fields using time-dependent density functional theory (TDDFT). Several factors that contribute to selection rules are revealed. Extending the targets to stereoscopic molecules, it is shown that the symmetries contributing to selection rules for molecules should be judged by the structural projections in laser polarization plane. This feature originates from the fact that the effective symmetries of molecules are dependent on the invariance of field-free Hamiltonian under the transformations involving rotation and reflection. For laser fields, the symmetries contributing to selection rules can be judged by Lissajous figures and their dynamical directivities (i.e., the temporal evolutions of the electric field vectors when they trace the Lissajous figures), which can be explained by the analysis of the time-dependent Hamiltonian. According to the obtained results, we present a practical approach to predict the allowed harmonic orders, which can be extended to more complex molecules and various forms of laser fields.

The paper is organized as follows. In Sec.\ \ref{model}, we describe the numerical method and laser parameters used in our simulations. In Sec.\ \ref{ars}, we demonstrate the selection rules based on the associated rotational symmetry (ARS) of the target-laser system. In Sec.\ \ref{alig}, we show that the symmetries contributing to selection rules for targets are dependent on the projections of targets by extending the targets from planar molecules to stereoscopic molecules. In Sec.\ \ref{field}, the symmetries of laser fields are discussed using orthogonal two-color (OTC) laser fields. In Sec.\ \ref{conclusion}, we present a summary of the work.

\begin{figure}[htb] \label{fig1}
\centerline{
\includegraphics[width=6cm]{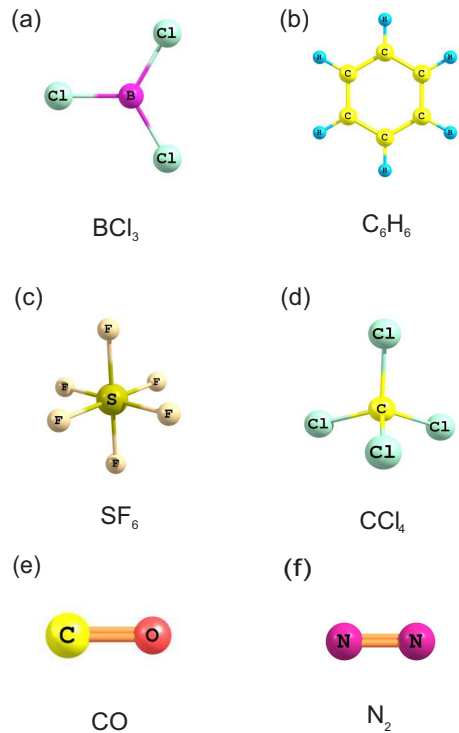}}
\caption{Illustrations of the structures of adopted molecules in this paper: (a) $\rm BCl_3$, (b) $\rm C_6H_6$, (c) $\rm SF_6$, (d) $\rm CCl_4$, (e) $\rm CO$, (f) $\rm N_2$. }
\end{figure}

\begin{figure*}[htb] \label{fig2}
\centerline{
\includegraphics[width=16cm]{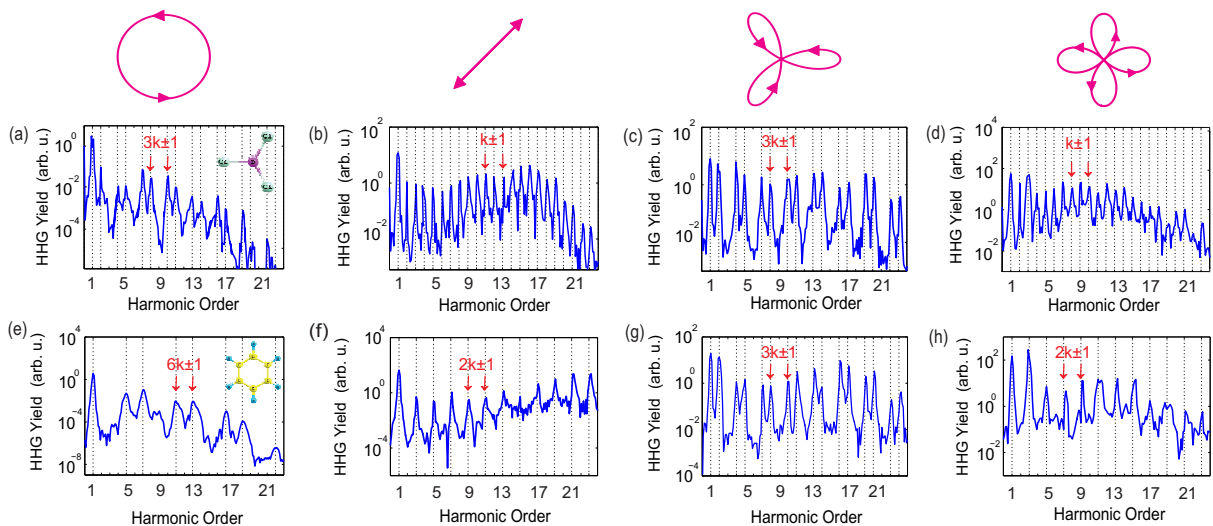}}
\caption{Harmonic spectra from (a)-(d) $\rm BCl_3 $ molecule and (e)-(h) $\rm C_6H_6 $ molecule with LP, CP, 1:2 CRB and 1:3 CRB laser fields. The Lissajous figures of laser fields are shown on the top row.}
\end{figure*}

\noindent \section{THEORETICAL MODEL} \label{model}
We numerically calculate the harmonic spectra from various targets in strong laser fields using the three-dimensional (3D) TDDFT  \cite{Runge}. In TDDFT method, the evolution of the system is described by a series of one-particle Kohn-Sham orbitals. Neglecting electron spin effects, the Kohn-Sham orbitals satisfy the time-dependent Kohn-Sham (TDKS) equations (atomic units are used throughout this paper unless otherwise stated)
\begin{gather} \label{TDKS}
i\frac{\partial}{\partial t}\psi_i(\mathbf{r},t) = [-\frac{\nabla ^2}{2}+v_{\rm eff}(\mathbf{r},t)]\psi_i(\mathbf{r},t),\ (i=1,2,\cdots,N).
\end{gather}
In Eq.(\ref{TDKS}), $ N $ is the number of Kohn-Sham orbitals $ \psi_i(\mathbf{r},t) $.
$ v_{\rm eff}(\mathbf{r},t) $ is the time-dependent Kohn-Sham potential and defined as
\begin{eqnarray} \label{VKS}
v_{\rm eff}(\mathbf{r},t) = v_{\rm H}(\mathbf{r},t)+v_{\rm xc}(\mathbf{r},t)+v_{ne}(\mathbf{r},t)+\mathbf{r}\cdot\mathbf{E}(t),
\end{eqnarray}
where $v_{\rm H}(\mathbf{r},t)$ is the Hartree potential, given by
\begin{eqnarray} \label{VH}
v_{\rm H}(\mathbf{r},t) = \int \frac{n(\mathbf{r'},t)}{|\mathbf{r}-\mathbf{r'}|}d\mathbf{r'}.
\end{eqnarray}
$n(\mathbf{r},t)$ is the time-dependent electron density written as $ n(\mathbf{r},t)=\sum_{i=1}^N|\psi_i(\mathbf{r},t)|^2 $. The Hartree potential accounts for the classical Coulomb interaction among the electrons. $ v_{\rm xc}(\mathbf{r},t) $ is the exchange-correlation potential, which includes all non-trivial many body effects. The exchange and correlation functional we use here are general gradient approximation (GGA) in the parametrization of Perdew-Burke-Ernzerhof (PBE) \cite{PBE}. $ v_{ne}(\mathbf{r},t) $ represents electron-ion interactions described with norm-conserving Troullier-Martins pseudopotentials \cite{TM} in the Kleinman-Bylander form \cite{KB}. $ \mathbf{E}(t) $ is the electric field of the laser pulse. The TDKS equations are discretized and solved with Octopus package \cite{octopus1,octopus2,octopus3}. In our numerical simulations, all of laser pulses are polarized in the $ x-y $ plane. The LP laser field is polarized along $x-$axis. We adopt a trapezoidal envelope with a total duration of 8 optical cycles (with 2-cycle linear ramps and 4-cycle constant center). The wavelength of the fundamental field is 800nm, and the intensity is $ 1\times10^{14}$ W/cm$^2$. The CRB laser field reads
\begin{eqnarray} \label{E_CRB}
\begin{aligned}
\mathbf{E}_{\rm {CRB}}(t) = &E_0f(t)\{[\textrm{cos}(\omega_0 t)+\textrm{cos}(q\omega_0 t)]\hat{\mathbf{e}}_x \\&+[\textrm{sin}(\omega_0 t)-\textrm{sin}(q\omega_0 t)]\hat{\mathbf{e}}_y\},
\end{aligned}
\end{eqnarray}
where $E_0$ is the field amplitude, and $f(t)$ is the envelope. $\omega_0$ is the fundamental frequency. $q$ is an integer greater than 1, which represent the frequency ratio of two CP components. $\hat{\mathbf{e}}_x$ and $\hat{\mathbf{e}}_y$ are the unit vectors in the $x-$ and $y-$directions respectively. The OTC laser pulses are composed of two mutually orthogonal laser fields with different frequencies. The OTC laser fields are described by \cite{ES2}
\begin{eqnarray} \label{E_OTC}
\mathbf{E}_{\rm OTC}(t) = E_0f(t)[\textrm{cos}(\omega_0 t)\hat{\mathbf{e}}_x+\textrm{cos}(q\omega_0 t+\varphi)\hat{\mathbf{e}}_y],
\end{eqnarray}
where $q$ is the frequency ratio of $x-$ and $y-$components. $\varphi$ is the relative phase of $x-$ and $y-$components, which is $\pi/2$ in our calculations. For both CRB and OTC laser fields, the intensity ratio of the two component laser fields is 1:1. The symmetries of laser fields remain unchanged when the intensities of the two laser field components are not equal (but still comparable), so the selection rules are the same as those with intensity ratio 1:1.

We apply the dipole approximation, which is used commonly for HHG. The harmonic spectrum is obtained by calculating the Fourier transform of the dipole acceleration \cite{Acceleration}
\begin{eqnarray} \label{Hw}
S(\omega) = \left|\int{\ddot{\mathbf{d}}(t)e^{i\omega t}dt}\right| ^2,
\end{eqnarray}
where $ \mathbf{d}(t) $ is the time-dependent dipole moment given by
\begin{eqnarray} \label{dt}
\mathbf{d}(t) =  \int{n(\mathbf{r},t) \mathbf{r}d\mathbf{r}}.
\end{eqnarray}

To reveal the selection rules of HHG, various kinds of molecules with different structures are adopted in our calculations as summarized in Fig.\ 1.  All the molecules lie in the laser polarization plane ($ x-y $ plane) except stereoscopic molecules shown in Figs.\ 1(c) and 1(d). The orientation effects of stereoscopic molecules will be studied in detail in Sec.\ \ref{alig}.

\noindent \section{associated rotational symmetries of molecule-laser configurations} \label{ars}
The selection rules of harmonic spectra originate from symmetries of systems. Based on Floquet formalism, the probability to get the $n$th harmonic in state $ \Psi_\varepsilon(\mathbf{r},t) $ is \cite{Circ1}
\begin{eqnarray} \label{n4}
\sigma_\varepsilon^{(n)} \propto n^4\left|\langle\langle\Phi_\varepsilon(\mathbf{r},t)|\hat{u}e^{-in\omega t}|\Phi_\varepsilon(\mathbf{r},t)\rangle\rangle\right|^2,
\end{eqnarray}
where $\Phi_\varepsilon(\mathbf{r},t)$ is given by $ \Psi_\varepsilon(\mathbf{r},t)=\Phi_\varepsilon(\mathbf{r},t)e^{-i\varepsilon t} $. The $\Phi_\varepsilon(\mathbf{r},t)$ is known as single Floquet state, which is the simultaneous eigenfunction of Floquet Hamiltonian $ \mathcal{\hat{H}}(t)=\hat{H}(t)-i\hbar\frac{\partial }{\partial t}$. $\varepsilon$ is called quasi-energy. $\hat{u}$ is the dipole moment operator. $\omega$ is the circular frequency of fundamental frequency field. Double bracket denotes the integral over space and time. The $n$th harmonic is emitted only if $ \sigma_\varepsilon^{(n)}\neq0 $. For planar systems with laser fields polarized in molecule plane, if the Floquet Hamiltonian $ \mathcal{\hat{H}}(t) $ is invariant under an $N$-fold transformation
\begin{eqnarray} \label{PN}
\hat{P}_N = (\varphi\rightarrow\varphi+\frac{2\pi}{N},t\rightarrow t+\frac{2\pi}{N\omega}),
\end{eqnarray}
where $\varphi$ is the azimuth angle and $t$ is the time, the nonzero term of $ \sigma_\varepsilon^{(n)} $ in Eq. (\ref{n4}) satisfies
\begin{eqnarray} \label{exp}
\mathrm{exp}[-i\frac{2\pi (n\pm 1)}{N}]=1.
\end{eqnarray}
Eq. (\ref{exp}) indicates that the allowed harmonics are $ n=kN\pm 1 $ orders, where $k$ is an integer. The result attributes selection rules to symmetry of Floquet Hamiltonian $ \mathcal{\hat{H}}(t) $ under rotation operator $ \hat{P}_N $. It is worth noting that the single electronic orbital does not always possess the same symmetry as the field-free Hamiltonian of system. However, the generation of the harmonics is contributed by a series of (degenerated) electronic orbitals. The total electron density of the electronic orbitals must possess the same symmetry as the field-free Hamiltonian, and the selection rules are dependent on the symmetries of the density distributions. Some numerical simulations were used to confirm the rules, but all of them were based on low-dimensional model planar molecules and/or SAE approximation.

Although the selection rules of HHG for planar molecules are derived based on above deduction. The allowed harmonic orders can be more intuitively judged by analyzing the symmetries of the target and laser field separately. If a molecule exhibits an invariance under a rotation of $2\pi/M$ ($M$ is a positive integer) around axis of laser propagation, this molecule possesses $M$-fold rotational symmetry, which is denoted as $C_M$. For example, when we rotate the $\rm BCl_3 $ molecule shown in Fig.\ 1(a) around the $z$ axis (axis of laser propagation) by $2\pi/3$, the configuration of the $\rm BCl_3 $ molecule remains the same. Therefore, the $ \rm BCl_3 $ molecule possesses $C_3$ symmetry. Likewise, the $\rm C_6H_6 $ molecule (shown in Fig.\ 1(b)) possesses $C_6$ symmetry. Essentially, the symmetry of a molecule is the reflection of invariance of field-free Hamiltonian under the transformation operator $\hat{P}_N$. For a laser field, the symmetry contributing to the selection rules is determined by the invariance of interaction term of full Hamiltonian under the $L-$fold transformation, which will be discussed in detail in Sec.\ \ref{field}. Analogous to molecules, the $L$-fold symmetries of laser fields are denoted as $C_L$. The Lissajous figures of the commonly discussed CP, LP, (1:2 and 1:3) CRB laser fields are presented in the top row of Figs.\ 2(a)-2(d), respectively. The LP laser field possesses $C_2$ symmetry and CP laser field possesses $C_\infty$ symmetry. For CRB laser fields with frequency ratio $q$, the Lissajous figure resembles a multiblade fan with $q+1$ lobes, and the CRB laser fields possess $ C_{q+1} $ symmetry. Therefore, the 1:2 CRB and 1:3 CRB possess $C_3$ and $C_4$ symmetry respectively. One can see that the symmetries contributing to the selection rules for laser fields are the same as rotational symmetries of Lissajous figures for these forms of laser fields.

The harmonic spectra from $ \rm BCl_3 $ and $ \rm C_6H_6 $ molecules driven by CP, LP, 1:2 CRB and 1:3 CRB laser fields are presented in Figs.\ 2(a)-2(h). Figures 2(a) and 2(e) show the harmonic spectra driven by CP laser fields. In Fig.\ 2(a), the allowed harmonic orders are $ 3k\pm1 $, which corresponds to the fact that the $ \rm BCl_3 $ molecule possesses $C_3$ symmetry. Similarly, in Fig.\ 2(e), the allowed harmonic orders are $ 6k\pm1 $, which corresponds to the $C_6$ symmetry of $ \rm C_6H_6 $ molecule. It is shown that the allowed harmonic orders driven by CP laser field are only determined by the symmetries of the molecular structures as $ kM\pm1 $. This is in agreement with the results in Ref. \cite{Circ1,Circ2,Circ3}.

Figures.\ 2(c) and 2(d) show harmonic spectra from $\rm BCl_3 $ driven by 1:2 and 1:3 CRB laser fields. The allowed harmonics are determined by the symmetries of molecules and laser fields according to a GCD rule as demonstrated in Ref. \cite{Mauger}. In Fig.\ 2(c), both the $ \rm BCl_3 $ molecule and 1:2 CRB laser field possess $C_3$ symmetry ($M=3$ and $L=3$). Since the GCD of $M$ and $L$ is 3, the target-laser system possesses an overall $C_3$ symmetry. Correspondingly, the allowed harmonic orders are $ 3k\pm1 $. In Fig.\ 2(d), the $ \rm BCl_3 $ molecule and 1:3 CRB laser pulse possess $C_3$ ($M=3$) and $C_4$ ($L=4$) symmetry, respectively. The GCD of $M$ and $L$ is 1, and thus the $ k\pm1 $ order harmonics are allowed. The harmonic spectra from $ \rm C_6H_6 $ molecule driven by 1:2 and 1:3 CRB laser pulse are presented in Figs.\ 2(g) and 2(h). In the same way, since the GCD of symmetries are 3 and 2, the allowed harmonic orders are $ 3k\pm1 $ and $ 2k\pm1 $, respectively.

Figures 2(b) and 2(f) show the harmonic spectra of $ \rm BCl_3 $ and $ \rm C_6H_6 $ molecules driven by LP laser pulse. The allowed harmonic orders are $ k\pm1 $ and $ 2k\pm1 $, respectively. The results can also be explained based on the GCD rule of symmetries: the LP laser field possesses $C_2$ symmetry, and therefore the GCDs of rotational symmetries are 1 for $ \rm BCl_3 $ and 2 for $ \rm C_6H_6 $.

The above results indicate that the selection rules of HHG with various kinds of targets and laser fields can be summarized according to the symmetries: if the target and laser field possess $M$-fold and $L$-fold symmetries, the allowed harmonic orders should be $ kN\pm 1$, where $N$ is GCD of $M$ and $L$. We refer to the $N$-fold symmetry of the target-laser system as ARS. When the laser field is CP, the ARS of target-laser system is the same as the $C_M$ symmetry of the target. Therefore, the selection rules are only determined by the symmetries of targets as shown in Figs.\ 2(a) and 2(e). For atomic target, since the target possesses $C_\infty$ symmetry, the ARS of target-laser system is the same as the $C_L$ symmetry of laser field. Therefore, the allowed harmonic orders depend only on the symmetry of laser field \cite{atom-sr}. As a special case, atomic targets do not radiate harmonics driven by CP laser fields, because the ARS is $C_\infty$. This selection rules have been confirmed by a number of our other numerical calculations. The physical origin of ARS dependent selection rules is the symmetry of full time-dependent Hamiltonian: the $C_N$ symmetry is exactly corresponding to the invariance of full Hamiltonian under the transformation $P_N$, while the $C_M$ and $C_L$ symmetry are only responsible for the field-free Hamiltonian and interaction term respectively. Nevertheless, it is more practical and intuitive to judge the allowed harmonic orders according to the ARS approach. In the following, we will show that the symmetry contributing to the ARS should be identified in a more general way for the target and the laser field.

\noindent \section{The selection rules for stereoscopic targets} \label{alig}

\begin{figure}[htb]
\centerline{
\includegraphics[width=8cm]{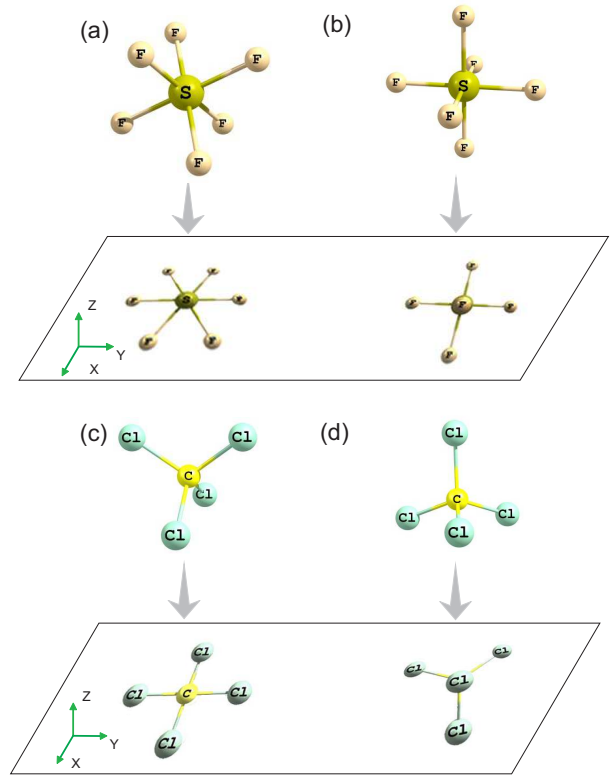}}
\caption{Illustrations of two different orientations for (a),(b) $\rm SF_6$ molecule and (c),(d) $\rm CCl_4$ molecule. The corresponding projections in $ x-y $ plane are shown at the bottom of figures.}
\end{figure}

In this section, the identification of symmetries contributing to the ARS for the targets will be discussed with stereoscopic molecules. We take the $\rm SF_6$ molecule as an example. We firstly consider the orientation of $\rm SF_6$ molecule shown in Fig.\ 3(a) (the top view from $z-$axis is shown in Fig.\ 4(a)). The harmonic spectra driven by CRB laser fields with frequency ratios 1:2 ($C_3$ symmetry), 1:3 ($C_4$ symmetry) and 1:5 ($C_6$ symmetry) are shown in Figs.\ 4(b)-4(d), respectively. One can see that allowed harmonic orders are $ 3k\pm 1 $, $ 2k\pm 1 $ and $ 6k\pm 1 $, respectively. As shown in Fig.\ 3(a), the $\rm SF_6$ molecule exhibits an invariance under a rotation of $2\pi/3$ around the $z$ axis, and therefore it possesses $C_3$ symmetry. If the symmetry of the target is considered as $C_3$, the obtained ARSs are $C_3$, $C_1$, $C_3$ and the allowed harmonic orders should be $ 3k\pm 1 $, $ k\pm 1 $ and $ 3k\pm 1 $, respectively. Obviously, these deduced selection rules conflict with the results found in Figs. 4(b)-4(d).

\begin{figure*}[htb]
\centerline{
\includegraphics[width=16cm]{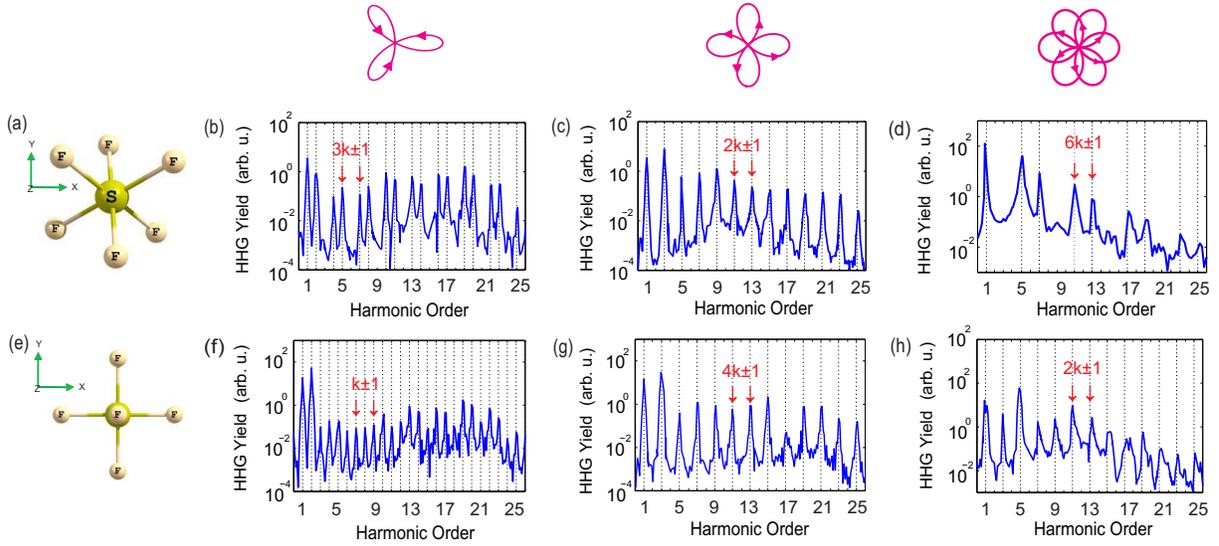}}
\caption{Harmonic spectra from (b-d) $\rm SF_6$ molecule orientated as in Figs.\ 3(a) and (f-h) $\rm SF_6$ molecule oriented as in Figs.\ 3(b) driven by CRB laser fields with frequency ratios 1:2, 1:3 and 1:5. The top views of these orientations from $z-$axis are shown in (a) and (e), respectively. The Lissajous figures of laser fields are shown on the top row.}
\end{figure*}

\begin{figure*}[htb]
\centerline{
\includegraphics[width=16cm]{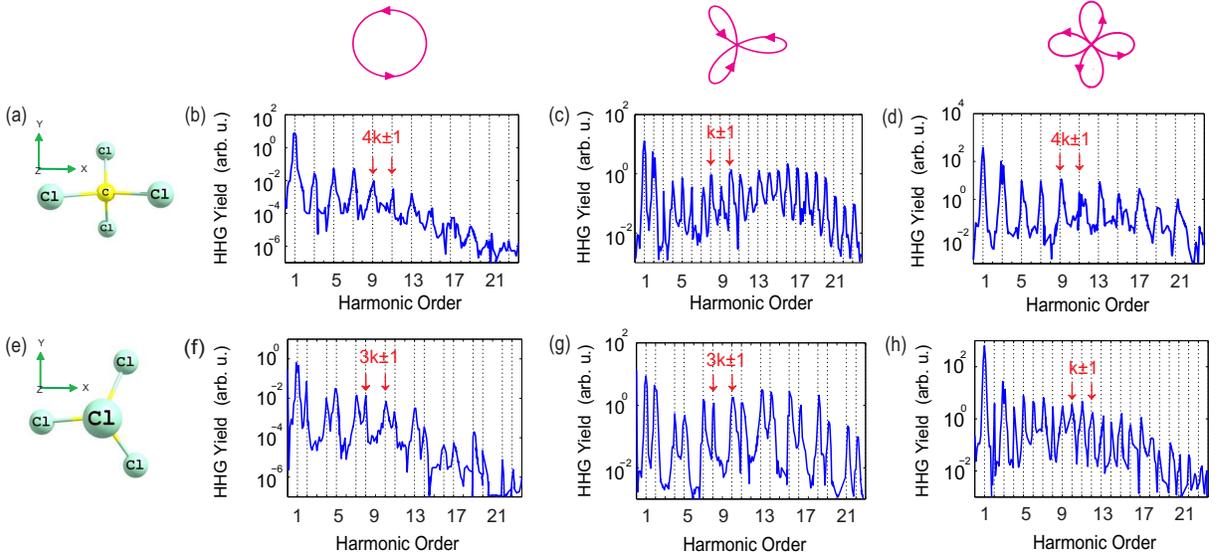}}
\caption{Harmonic spectra from (b-d) $\rm CCl_4$ molecule oriented as in Figs.\ 3(c) and (f-h) $\rm CCl_4$ molecule oriented as in Figs.\ 3(d) driven by CP, 1:2 CRB and 1:3 CRB laser fields. The top views of these orientations from $z-$axis are shown in (a) and (e), respectively. The Lissajous figures of laser fields are shown on the top row.}
\end{figure*}

The discrepancy between deduction and results originates from the improper identification of the symmetry contributing to ARS for targets. To correctly obtain selection rules, the effective symmetry of a target should be dependent on the rotational symmetry of its projection on the polarization plane, rather than the rotational symmetry of the target itself. For the orientated $\rm SF_6$ molecule in Fig.\ 3(a), the projection on the polarization $ x-y $ plane forms a regular hexagon with $C_6$ symmetry. Thus, the ARSs in 1:2, 1:3, 1:5 CRB laser fields are $C_3$, $C_2$, $C_6$ and the allowed harmonic orders should be $ 3k\pm 1 $, $ 2k\pm 1 $ and $ 6k\pm 1 $, respectively. These allowed harmonic orders deduced from the symmetries of projections agree with the obtained results in Fig.\ 4(b)-4(d). The $\rm SF_6$ molecule can alternatively be orientated as in Fig.\ 3(b). In this orientation, the projection on the $x$-$y$ plane forms a square possessing $C_4$ symmetry. Therefore, the allowed harmonic orders will be $ k\pm 1 $, $ 4k\pm 1 $ and $ 2k\pm 1 $, respectively. The calculated harmonic spectra are shown in Figs.\ 4(f)-4(h). It is shown that the allowed harmonic orders agree well with the prediction. Note that the observed harmonic orders are the same for the $ 2k\pm 1 $ rule and $ 4k\pm 1 $ rule. Therefore, the two selection rules can not be distinguished from the intensity spectrum. To clearly distinguish the two different selection rules, we further investigate the polarization properties of the harmonics. The study \cite{Circ1,Circ2,Mauger} shows that the $ 4k+1 $ and $ 4k-1 $ order harmonics (the ARS is $C_4$) are CP in opposite helicities when molecules interact with CP or CRB laser pulses, but the harmonics with $ 2k\pm 1 $ selection rule (the ARS is $C_2$) are arbitrarily polarized in CRB laser pulses. Figures\ 6(a) and 6(b) show the calculated ellipticities of HHG spectra corresponding to Figs.\ 4(g) and 4(h), respectively. In Fig.\ 6(a), the alternation of left-handed and right-handed circularly polarizations conforms the $ 4k\pm 1 $ selection rule in Fig.\ 4(g). On the contrary, in Fig.\ 6(b), the randomly varying ellipticities also conform the $ 2k\pm 1 $ selection rule in Fig.\ 4(h). For planar targets (such as atoms, $\rm BCl_3$, $\rm C_6H_6$, {\it etc.}) located in the polarization plane, the symmetries of the targets and the symmetries of their projections are exactly the same. In this case, the role of the projections for the selection rules could not be distinguished as in previous studies.

From the above discussion, it is shown that the effective symmetries of molecules are dependent on the projections of molecules. The dependence of projections of targets on selection rules can be understood by the three-step picture of HHG. In the HHG process, the wavepacket of the continuum state propagates in the $x-y$ plane and returns to the parent core generating high harmonics.  According to this model, the generated harmonic emission is the same for the target and its mirror image in the $x-y$ plane, i.e., the field-free Hamiltonian $H_0(x,y,z)$ from $H_0(x,y,-z)$ can not be distinguished in HHG. This leads to the phenomenon that the effective symmetry contributing to ARS are determined by the projection of a molecule instead of itself. In essence, the dependence of projections of molecules on the selection rules originates from the symmetry of full time-dependent Hamiltonian as described in Sec.\ \ref{ars} for planar molecules. However, the theory should be generalized for stereoscopic molecules. For a stereoscopic system, the rotation transformation $(\varphi\rightarrow\varphi+\frac{2\pi}{N})$ should be substituted by the transformation $(\mathbf{r} \rightarrow \hat{O}_{N}(\mathbf{r}))$, where $\hat{O}_{N}$ is the geometric operation of a rotation around $z-$axis with or without an accompanying reflection in the $z=0$ plane. The $N-$fold transformation operator $\hat{O}_{N}$ is written as
\begin{eqnarray} \label{ON}
\hat{O}_{N} = (\varphi\rightarrow\varphi+\frac{2\pi}{N},z\rightarrow \pm z).
\end{eqnarray}
In conclusion, the selection rules are $ kN\pm 1$ when the full time-dependent Hamiltonian is invariant under a combined transformation $(\mathbf{r} \rightarrow \hat{O}_{N}(\mathbf{r});t\rightarrow t+\frac{2\pi}{N\omega})$. The field-free Hamiltonian is only involved in the $\hat{O}_{N}$ operation because of its independence of time, so the $M-$fold symmetry contributing to the selection rules for molecules should be defined by the invariance of field-free Hamiltonian under transformation $\hat{O}_{M}$. Therefore, the reflection transformation of field-free Hamiltonian results in the fact that the symmetry contributing to the ARS is dependent on the symmetry of projection of a molecule rather than symmetry of the molecule itself. For example, the configuration of $\rm SF_6$ molecule shown in Fig.\ 3(a) is invariant under a rotation by $2\pi/6$ and a reflection in the $x-y$ plane. Therefore, the $\hat{O}_{M}$ is expressed as $(\varphi\rightarrow\varphi+\frac{2\pi}{6},z\rightarrow -z)$, and the effective symmetry of such molecule is $C_6$. Similarly, the configuration of $\rm SF_6$ molecule shown in Fig.\ 3(b) is invariant under the transformation $(\varphi\rightarrow\varphi+\frac{2\pi}{4},z\rightarrow z)$, which results in the $C_4$ symmetry in this orientation. Although the symmetry contributing to the selection rules for a molecule is essentially determined by the invariance of full time-dependent Hamiltonian under the transformation $\hat{O}_{M}$, it can be more intuitively judged by the projection.

The dependence of the projection on selection rules demonstrates an additional characteristic for the selection rules: the allowed harmonics are sensitive to molecular orientations. This is because the projections possess different rotational symmetries when the same molecules are oriented in different directions. Consequently, the ARSs are different corresponding to different orientations. For the same molecule, when the molecular orientation changes, the allowed harmonics change with it.

\begin{figure}[htb]
\centerline{
\includegraphics[width=8cm]{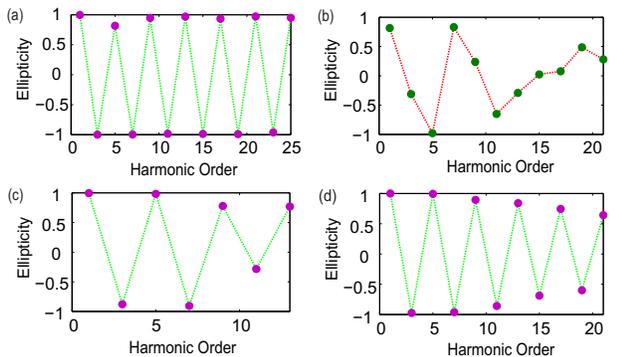}}
\caption{The ellipticities of HHG spectra below the range of cutoff corresponding to Fig.\ 4(g), Fig.\ 4(h), Fig.\ 5(b) and Fig.\ 5(d), respectively. The purple dots represent the $ 4k\pm 1 $ orders, and the green dots represent the $ 2k\pm 1 $ orders.}
\end{figure}

The $\rm CCl_4$ molecule is also adopted to demonstrate the dependence on projections of molecules and the orientation dependence of selection rules. Two orientations for $\rm CCl_4$ molecule are considered. When the $\rm CCl_4$ molecule is oriented as in Fig.\ 3(c) (the top view from $z-$axis is shown in Fig.\ 5(a)), its projection possesses $C_4$ symmetry (the molecule possesses $C_2$ symmetry). The calculated harmonic spectra driven by CP and CRB laser pulses with frequency ratios 1:2 ($C_3$ symmetry) and 1:3 ($C_4$ symmetry) are presented in Figs.\ 5(b)-5(d), respectively. According to the above discussions, the allowed harmonic orders should be $ 4k\pm 1 $, $ k\pm 1 $ and $ 4k\pm 1 $, which is consistent with the results in Figs.\ 5(b)-5(d). In order to distinguish the $4k\pm 1$ rule from $2k\pm1$ rule, the ellipticities of allowed harmonics in Figs.\ 5(b) and 5(d) are shown in Figs.\ 6(c) and 6(d), respectively. The alternation of helicities of the (nearly) circularly polarizations confirms the $4k\pm 1$ rule in Figs.\ 5(b) and 5(d). The other orientation is shown in Fig.\ 3(d) (the top view from $z-$axis is shown in Fig.\ 5(e)). It is found that, when the $\rm CCl_4$ molecule is rotated to the other orientation, the generated harmonics are significantly changed in the same laser field. This is because the symmetry of the projection changes from $C_4$ to $C_3$, and then ARSs change from $C_4$, $C_1$, $C_4$ to $C_3$, $C_3$, $C_1$ respectively. As a result, the allowed harmonic orders are $ 3k\pm 1 $, $ 3k\pm 1 $ and $ k\pm 1 $ as shown in Figs.\ 5(f)-5(h).

Our results imply that allowed harmonics not only contain fingerprints of molecular structures, but also reveal the information of orientations. Currently, there are various approaches to probe the symmetry of electronic orbitals using high harmonic spectroscopy or strong-field photoelectron spectrum \cite{prosym1,prosym2,prosym3,prosym4,prosym5}. By comparison, only a few works pay attention to decode symmetries of molecular geometric structures with HHG, especially for stereoscopic molecules. Here, the selection rules provide a feasible scheme. The three-dimensional structures can be decoded from the harmonic spectra at different orientations according to the allowed harmonic orders. On the other hand, molecular orientations can be evaluated according to allowed harmonics. For instance, this idea has been used to check the orientation of linear molecules \cite{Baykusheva}. This method will show greater advantages for stereoscopic molecules.

\noindent \section{the symmetries of laser fields} \label{field}

Besides the target, the ARS is also dependent on the symmetry of the laser field. In the previous calculations, it was found that the symmetries of laser fields can be directly judged by the rotational symmetries of their Lissajous figures intuitively. For example, the configuration of a 1:2 CRB laser field remains the same under a rotation by $2\pi/3$ in polarization plane, which corresponds to the $C_3$ symmetry of the laser field. Is it general that the symmetry of a laser field can be intuitively judged only by the geometric structure? To discuss on the question, we adopt the OTC laser fields.

\begin{figure}[htb]
\centerline{
\includegraphics[width=6cm]{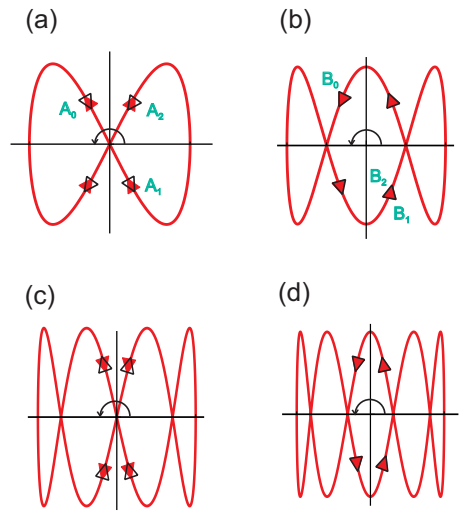}}
\caption{The Lissajous figures of OTC laser fields with frequency ratios (a) 1:2, (b) 1:3, (c) 1:4 and (d) 1:5. The red filled arrows indicate the rotation directions of laser fields, and the black hollow arrows indicate rotation directions after the laser fields are rotated by ${2\pi/2}$. $\mathbf A_0$, $\mathbf A_1$, $\mathbf A_2$, $\mathbf B_0$, $\mathbf B_1$ and $\mathbf B_2$ denote the electric vectors corresponding to the red filled (or black hollow) arrows.}
\end{figure}

\begin{figure}[htb]
\centerline{
\includegraphics[width=8cm]{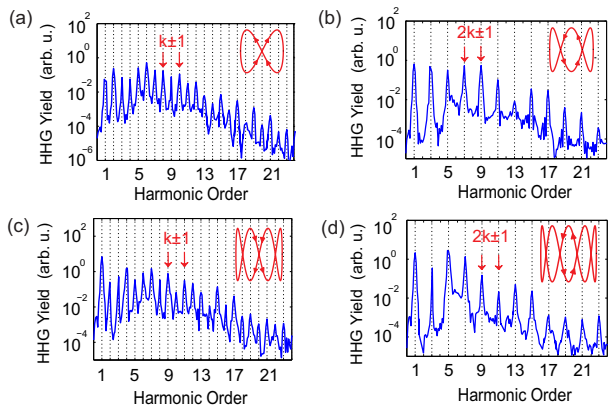}}
\caption{The harmonic spectra from H atom driven by OTC laser pulses with frequency ratios (a) 1:2, (b) 1:3, (c) 1:4 and (d) 1:5. The Lissajous figures of the lasers fields are plotted in the insets.}
\end{figure}

In Figs.\ 7(a)-7(d), the Lissajous figures of OTC laser fields with frequency ratios 1:2, 1:3, 1:4 and 1:5 are presented. If only the geometric structure of laser field is considered, one would conclude that all of the OTC laser fields in Figs.\ 7(a)-7(d) possess the same  $C_2$ symmetry. To examine the actual symmetry contributing to the ARS, the harmonic spectra from H atom driven by the four OTC laser fields are obtained as shown in Figs.\ 8(a)-8(d). The H atom is employed because the ARS is exactly the symmetry of the laser field for atomic target. It is found that the allowed harmonic orders are not the same with the four laser fields. When the frequency ratios are 1:2 and 1:4, the allowed harmonic orders are $k\pm1$ (Figs.\ 8(a) and 8(c)), which indicates the laser fields possess $C_1$ symmetry. When the frequency ratios are 1:3 and 1:5, the allowed harmonic orders are $2k\pm1$ (Figs.\ 8(b) and 8(d)), indicating that the laser fields possess $C_2$ symmetry. These results definitely show that symmetries of laser fields can not be intuitively judged by only the geometric structures.

Comparing the laser fields with even frequency ratios (1:2 and 1:4) and odd frequency ratios (1:3 and 1:5), it is found that the symmetries are also dependent on the temporal evolutions of electric field vectors when they trace the Lissajous figures. The temporal evolutions of electric field vectors are called dynamical directivities of laser fields. In Figs.\ 7(a) and 7(c), the red filled arrows indicate the rotation directions of laser fields, and the black hollow arrows indicate rotation directions after laser fields are rotated by ${2\pi/2}$. Note that the red and black arrows do not coincide. This means the laser fields do not exhibit an invariance under a rotation of ${2\pi/2}$ considering the dynamical directivities. Therefore, the symmetries of the laser fields are $C_1$ instead of $C_2$ respectively, and the allowed harmonic orders are $k\pm1$ instead of $2k\pm1$ as shown in Figs.\ 8(a) and 8(c). For laser fields shown in Figs.\ 7(b) and 7(d), the coincidence of red and black arrows shows that symmetries of the laser fields are still $C_2$ when dynamical directivities are considered, and thus the allowed harmonic orders are still $2k\pm1$ as shown in Figs.\ 8(b) and 8(d). Dating back to LP, CP and CRB laser fields used in Sec.\ \ref{ars} and Sec.\ \ref{alig}, the symmetries do not change when dynamical directivities of laser fields are taken into account. For example, when 1:2 CRB laser field is rotated by ${2\pi/3}$, the dynamical directivity also remains the same.

\begin{figure}[htb]
\centerline{
\includegraphics[width=8cm]{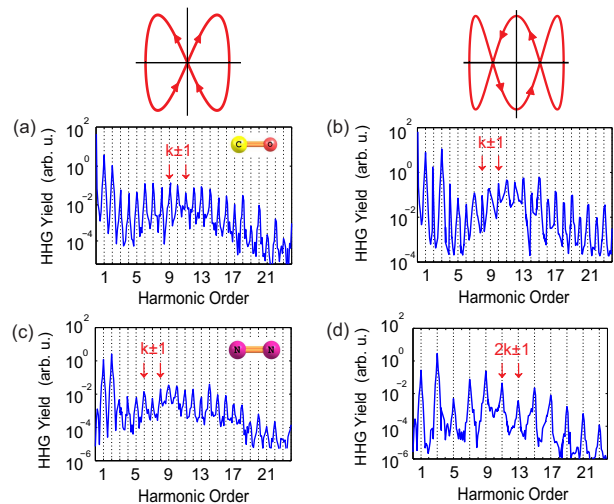}}
\caption{Harmonic spectra from (a),(b) $\rm CO $ and (c),(d) $\rm N_2 $ driven by OTC laser pulses with frequency ratios 1:2 and 1:3. The Lissajous figures of laser fields are shown on the top row.}
\end{figure}

The dependence of the symmetry on dynamical directivity of a laser field should be explained by the symmetry of full Hamiltonian. The selection rules of HHG are determined by the invariance of full time-dependent Hamiltonian under a combined transformation $(\mathbf{r} \rightarrow \hat{O}_{N}(\mathbf{r});t\rightarrow t+\frac{2\pi}{N\omega})$. For the interaction with a laser field in the dipole approximation, the interaction term is $\mathbf{r}\cdot \mathbf{E}(t)$. In the laser polarization plane, the interaction term is always invariant under the reflection transformation because the laser field has no projection onto the $z-$axis. In this case, the symmetry of a laser field only needs to be judged by rotation transformation. When the radial vector $\mathbf{r}$ is rotated under the operator $\hat{O}_{N}$, the scalar product $\mathbf{r}\cdot \mathbf{E}(t)$ remains invariant only if the time transformation $(t\rightarrow t+\frac{2\pi}{N\omega})$ results in the same rotation of electric vector $\mathbf{E}(t)$. Therefore, the $L-$fold symmetry of a laser field is defined by
\begin{eqnarray} \label{P_E}
\mathbf{E}(t+\frac{2\pi}{L\omega}) = \hat{O}_{L}(\mathbf{E}(t)).
\end{eqnarray}
The symmetries of all aforementioned laser fields (such as the $C_3$ symmetry of 1:2 CRB laser field) are determined by Eq. (\ref{P_E}). In Fig.\ 7(a), the electric vector at $\mathbf A_0$ is transformed to $\mathbf A_2$ under the time transformation $(t\rightarrow t+\frac{2\pi}{2\omega})$, while the $\mathbf A_0$ is transformed to $\mathbf A_1$ by the rotation transformation $(\varphi\rightarrow\varphi+\frac{2\pi}{2})$. The $\mathbf A_1$ and $\mathbf A_2$ do not coincide, so the symmetry of the laser field is not $C_2$. By comparison, in Fig.\ 7(b), the electric vector at $\mathbf B_0$ is changed to $\mathbf B_1$ by a rotation transformation $(\varphi\rightarrow\varphi+\frac{2\pi}{2})$ and is changed to $\mathbf B_2$ by a time transformation $(t\rightarrow t+\frac{2\pi}{2\omega})$, respectively. The coincidence of $\mathbf B_1$ and $\mathbf B_2$ reveals the $ C_2$ symmetry of the laser field. The symmetry of a laser field is essentially determined by the symmetry of interaction term of full Hamiltonian. However, it can be judged according to the symmetry of the geometric structure and dynamical directivity intuitively.

In Figs.\ 9(a)-9(d), the molecular targets $\rm CO$ and $\rm N_2$ are adopted to further demonstrate the dependence of the selection rules on the dynamical directivities of laser fields. In Figs.\ 9(a) and 9(b), harmonic spectra from $\rm CO$ molecule driven by OTC laser fields with frequency ratios 1:2 and 1:3 are presented, respectively. It is shown that the allowed harmonic orders are $k\pm1$ in both cases. This is because $\rm CO$ molecule possesses $C_1$ symmetry and thus the ARS are always $C_1$ regardless of field fields. Harmonic spectra from $\rm N_2$ molecule driven by OTC laser fields with frequency ratios 1:2 and 1:3 are shown in Figs.\ 9(c) and 9(d), respectively. For 1:2 OTC laser field, the allowed harmonic orders are $k\pm1$, because the ARS of the system with $\rm N_2$ molecule ($C_2$ symmetry) and 1:2 OTC field ($C_1$ symmetry) is $C_1$. For 1:3 OTC laser field, the allowed harmonic orders are $2k\pm1$, because both $\rm N_2$ molecule and 1:3 OTC laser field possess $C_2$ symmetry (the ARS is $C_2$). Our calculations show that the dynamical directivity of the laser field is an important aspect for the judgment of symmetry contributing to ARS. The symmetries of the laser fields can be identified according to both the geometrical structure and the dynamical directivity intuitively.

\noindent \section{Conclusion} \label{conclusion}
In summary, the selection rules of HHG are investigated with various real molecules and laser fields using TDDFT. The origin of the selection rules is discussed based on the symmetry of the full time-dependent through the paper. Moreover, it is shown that the selection rules can also be intuitively judged by the ARS of the target-laser configuration. Several factors that contribute to the ARS are revealed. For the stereoscopic target, we show that the ARS is contributed by the symmetry of the projection of the target rather than by the symmetry of the target itself. Correspondingly, it is shown that the allowed harmonics are dependent on the orientation of the target, which implies potential applications to probe the three-dimensional structure of the target molecule or to evaluate orientation. For the laser field, it is shown that the symmetry contributing to ARS can be judged by the symmetries of Lissajous figure and its dynamical directivity. In this work, we present a systematic study on the selection rules of HHG. From the results and discussions, a practical method to get selection rules is proposed, which can be extend to more complex molecules and various laser fields.

\section*{Acknowledgment}
This work was supported by the National Natural Science Foundation of China under Grants No. 11234004, No. 11404123 and No. 61275126, and the 973 Program of China under Grant No. 2011CB808103. Numerical simulations presented in this paper were carried out using the High Performance Computing Center experimental testbed in SCTS/CGCL (see http://grid.hust.edu.cn/hpcc).

\end{document}